\documentclass[a4paper, 12pt]{amsart}

\usepackage{amssymb}

\def\dim{\hbox{{\rm dim}}}              

\newtheorem{definition}{Definition}

\begin{document}

\title{The Geometry of integrable and superintegrable systems}

\author{A. Ibort, G. Marmo$^1$}

\address{Depto. de Matem\'aticas, Univ. Carlos III de Madrid, Avda. de la Universidad 30, 28911 Legan\'es, Madrid, Spain.}\footnote{On leave of absence from the Depto. di Fisica Teorica, Univ. Federico II di Napoli, Via Cintia, Napoli, Italia.}

\email{marmo@infni.na.it, albertoi@math.uc3m.es}

\maketitle

\begin{abstract}  The group of automorphisms of the geometry of an integrable system is considered.    The geometrical structure used to obtain it  is provided by a normal form representation of integrable systems that do not depend on any additional geometrical structure like symplectic, Poisson, etc.   Such geometrical structure provides a generalized toroidal bundle on the carrier space of the system.    Non--canonical diffeomorphisms of such structure generate alternative Hamiltonian structures for complete integrable Hamiltonian systems.   The energy-period theorem provides the first non--trivial obstruction for the equivalence of integrable systems. 

\end{abstract}


\section{Introduction}

After the Erlangen Programme written by Felix Klein with the collaboration of Sophus Lie (Lie visited Klein for two months, just before the programme was written) it is by now clear that a ``geometry'' or a ``geometrical structure'' on a manifold $M$ amounts to select a subgroup of the group of diffeomorphisms of $M$ \cite{Kl72}, \cite{Kl92} (a recent transcription is available on the arXiv: 0807.3161 v1[math.H0]).

In Physics we are quite familiar with this  correspondence, for instance we have the Poincar\'e group for the Minkowski space--time in Special Relativity, the group Diff$(\mathbb{R}^4)$ in general relativity,  the group of symplectomorphisms in Hamiltonian dynamics, contact transformations, the unitary group in Quantum Mechanics and so on.
In all previous cases, the subgroup of the group of diffeomorphisms determined by the geometry characterizes the geometry itself, for instance, if $(M_1, \omega_1)$ and $(M_2, \omega_2)$ are two symplectic manifolds and there is a group isomorphism $\Phi\colon \mathrm{Sp}(M_1,\omega_1) \to \mathrm{Sp}(M_2,\omega_2)$ of the corresponding groups of  canonical transformations, i.e., symplectic diffeomorphisms defined by each one, then the two symplectic manifolds are (up to a conformal constant) symplectically equivalent, i.e., there exists a diffeomorphisms $\varphi \colon M_1 \to M_2$ such that $\varphi^* \omega_2 = c \omega_1$ \cite{Ba86}-\cite{Ba88}.   Similar results were established by Grabowski for Poisson and Jacobi manifolds \cite{Gr00}.     Thus the autormorphisms determined by some geometrical structures are essentially inner as it happens for the group of unitary transformations of a Hilbert space.

In this paper we would like to identify the appropriate subgroup of Diff$(M)$ associated with integrable and superintegrable systems.

Thus if we have an integrable Hamiltonian dynamical system $\Gamma$ defined on a symplectic manifold $(M,\omega)$ the naive thought that the subgroup of the group of diffeomorphisms determined by it should be a subgroup of the group of canonical transformations of $\omega$ is immediately shown to be inadequate because integrable ssytems always admit alternative Hamiltonian descriptions and we would not know which canonical transformations to consider.

\medskip

To show how a geometrical structure is characterized by a tensorial object (and the associated invariance subgroup)  we consider a few examples.   We will consider first a linear structure on a manifold.   A possible linear structure on a manifold $M$ is characterized by a complete vector field $\Delta$ such that:

\begin{enumerate}
\item  There exists only one point, say $m_0 \in M$, such that $\Delta (m_0) = 0$.

\item  The eigenvalue problem:  $\mathcal{L}_\Delta f = 0\cdot f$, $f \in \mathcal{F}(M)$ has only trivial solutions, i..e, $f =$ constant if $M$ is connected.

\item  The eigenvalue problem:  $\mathcal{L}_\Delta f =  f$, $f \in \mathcal{F}(M)$ has $\dim M = n$ independent solutions $f_k$ such that $df_1\wedge \cdots \wedge df_n \neq 0$.

\end{enumerate}

Let us remark that the completeness condition on $\Delta$ allows to ``gene\-rate'' all of $M$ starting with a properly chosen transversal codimension one submanifold in a neighborhood of $m_0$ considered as a set of ``initial conditions''.    Any such vector field $\Delta$ identifies a subgroup of Diff$(M)$ by requiring $\varphi_*(\Delta) = \Delta$.   Thus the subgroup $GL(M,\Delta)$ of Diff$(M)$ of diffeomorphisms $\varphi$ preserving $\Delta$ is exactly the group of linear isomorphisms $GL(n,\mathbb{R})$ with $n= \dim M$ where we use the global chart provided by the functions $f_j$ to identify $M$ with $\mathbb{R}^n$.   In infinitesimal terms, linear vector fields are then solutions of  $[X,\Delta] = 0$.    Moreover if we have two linear structures $(M_1, \Delta_1)$ and $(M_2,\Delta_2)$ defined in two manifolds $M_1$ and $M_2$, and there is a group isomorphisms $\Psi \colon GL(M_1,\Delta_1) \to GL(M_2, \Delta_2)$, then there exists a {\em diffeomorphism} $\psi\colon M_1 \to M_2$ such that $\psi_*\Delta_1 = \Delta_2$.   

The notion of linear structure can be weakened by replacing conditions (1)-(3) above by asking that the set of points in $M$ satis\-fying $\Delta (m) = 0$ define a smooth submanifold $Z$ of $M$ of dimension $k$.  Condition (2) will be rephrased by asking that functions $f$ such that  $\mathcal{L}_\Delta f = 0$, defines an Abelian algebra whose spectrum is diffeomorphic to $Z$ and, finally, condition (3) will be substituted by demanding that $\mathcal{L}_\Delta f =  f$, define $n-k$ functionally independent fibrewise-linear functions.   Such ``partial'' linear structure is actually equivalent to a vector bundle structure on the manifold $M$ and the corresponding vector bundle autormophisms are selected by the condition $\varphi_*\Delta = \Delta$.

Another example in the same vein is concerned with possible tangent and cotangent bundle structures on a manifold $M$.   

We recall that a tangent bundle structure on $M$ is identified by a pair $(\Delta, S)$, where $\Delta$ defines a partial linear structure on $M$, i.e., a vector bundle structure, and $S$ is a $(1,1)$--tensor field (see for instance \cite{Fi89}) such that $\ker S = \mathrm{Im} \, S$ (that implies $S^2 = 0$), $S(\Delta ) = 0$ and $d_S^2 = 0$ where the ``twisted'' differential $d_S$ is defined by 
\begin{equation}\label{twisted_diff}
d_Sf (X) = df(S(x))  .
\end{equation}
   The vector field $\Delta$ is required to define a vector bundle structure on $M$, i.e., to satisfy the modified conditions (1), (2) and (3) above with $2k = \dim M$.   The vector bundle structure identified by functions in (3) becomes the tangent bundle structure of a manifold $Q$, so $M = TQ$.    Notice that the $(1,1)$--tensor $S$ is the soldering form of $TQ$.    In natural bundle coordinates $(q,v)$, these tensor fields take the form:
$$ S = dq \otimes \frac{\partial}{\partial v}, \quad \Delta = v\frac{\partial}{\partial v} .$$
The subgroup of Diff$(M)$ identified by $\varphi_*\Delta = \Delta$ and $\varphi_*S = S$ is the group of tangent bundle automorphisms. 
A single manifold can be equipped with alternative tangent bundle structures by considering diffeomorphisms $\varphi \colon M \to M$ such that $\varphi_* S = S$ but not preserving $\Delta$, we consider $\Delta_1$ and $\Delta_2$ given as $\Delta_1 = \Delta$ and $\Delta_2 = \varphi_*\Delta$.  Then $M$ becomes a double vector bundle if $[\Delta_1, \Delta_2] = 0$ and $M$ will carry two tangent bundle structures $TQ_1$ and $TQ_2$.  

Similarly one may define a cotangent bundle structure by means of a pair $(\Delta, \theta)$ where again, $\Delta$ is a partial linear structure on $M$, $\theta$ is a particular one--form such that $d\theta$ is a symplectic form, $i_\Delta d\theta = \theta$, and requiring that solutions of $\mathcal{L}_\Delta f =  0$ are pairwise in involution and define a maximal Abelian subalgebra  with respect to the the Poisson bracket defined by $\omega = d\theta$.  Notice that this is equivalent to asking that the submanifold $Q$ defined by $\Delta (m) = 0$ is Lagrangian with respect to $d\theta$.   Thus $M = T^*Q$ and the canonical Liouville one--form of $T^* Q$ is just the one--form $\theta$ above.
Again as in the case of tangent bundle structures, alternative cotangent bundle structures on $M$ can be constructed by choosing diffeomorphisms $\varphi$ such that $\varphi^* \theta \neq  \theta$.

Thus we will try to study the geometry of an integrable system $\Gamma$ by determining its associated subgroup of the group of diffeomorphisms of the manifold by writing the system in ``normal form'' instead of determing such a subgroup from the subgroup of diffeomorphisms determined by some geometrical structure determined by it.   In particular we shall consider systems with orbits possessing a compact closure, even though systems possessing unbounded orbits are relevant in scattering theory for instance.  

To indentify in such a manner the subgroups of diffeomorphisms corresponding to (super)integrable systems let us first recall the standard definition for integrable systems.

\section{The normal form of an integrable system}\label{integrability}

\subsection{Integrability and alternative Hamiltonian descriptions}
On a $2n$--dimensional symplectic manifold $(M,\omega)$ a vector field $\Gamma$ such that $i_\Gamma\omega = dH$ is said to be integrable if there exist $n$ functionally independent first integrals $f_1, \ldots, f_n$, $df_1 \wedge \cdots \wedge df_n \neq 0$, such that:
\begin{equation}\label{integrable}
\{ f_j, f_k \} = 0, \quad \{ H, f_k \} = 0, \quad \forall j,k = 1,2,\ldots, n.
\end{equation}
The independence condition $df_1 \wedge \cdots \wedge df_n \neq 0$ may just hold on some open dense submanifold of $M$, or in a weaker form, on some open invariant submanifold.  We will find this situation for instance when considering scattering problems were we prefer to remove closed regions composed of bounded orbits from phase space.

When the system $\Gamma$ possesses more than $n$ first integrals, the system is said to be superintegrable.  In this case, we have $f_1, \ldots, f_{n+k}$, $df_1 \wedge \cdots \wedge df_{n+k} \neq 0$, and $\{ H, f_j \} = 0$, for all $j = 1,2, \ldots, n+k$.  In the particular case of $n+k = 2n-1$ the system is said to be maximally superintegrable.   We remark that most of the known integrable systems are also maximally superintegrable.

To study the subgroup of diffeomorphisms of $M$ appropriate for an integrable system, it is convenient to have a ``normal form''.

As it was discussed in the introduction, in searching for normal forms, it is quite natural to ask which transformations are allowed on the sytem.  Thus for Hamiltonian systems, it would be natural to use canonical transformations, this is to consider transformations of the systems belonging to the closed subgroup of symplectic diffeomorphisms of the group of diffeomorphisms of our manifold $M$.  However, any integrable system admits alternative Hamiltonian descriptions, i.e., there are $(\omega_a,H_a)$, $a= 1,2$, such that 
$$ i_{\Gamma} \omega_1 = dH_1, \quad  i_{\Gamma} \omega_2 = dH_2 . $$
In this case, which canonical transformations should we use, symplectic diffeomorphisms with respect to $\omega_1$ or with respect to $\omega_2$?

In this respect it is amusing to recall a quarrel between Levi--Civita and Birkhoff around this.  Indeed in his 1934 paper ``A general survey of the theory of adiabatic invariants'' \cite{Le34} Levi--Civita feels the need to write a section entitled ``Birkhoff's severity against canonical variables and methods.  Apology for a milder attitude'' (p. 430).     From the point of view of Birkhoff, one should consider the orbit of vector fields obtained by acting on $\Gamma$ with Diff$(M)$.  All the vector fields in the same orbit share the same properties.  Thus, it makes sense to restrict the attention to just one representative.  Let us elaborate on this point in the particular case of linear systems.

We notice that a Hamiltonian vector field admits a factorization in terms of a Poisson tensor and an exact one--form:
\begin{equation}\label{factorization1} 
\Gamma = \Lambda (dH) .
\end{equation}
Clearly, any vector field in the orbit of $\Gamma$ will also be decomposable as above.  Moreover,
 if we consider diffeomorphisms $\varphi \in \mathrm{Diff}(M)$ such that $\varphi_* \Gamma = \Gamma$,
and apply it to the decomposition above (\ref{factorization1}), we find:
$$ \Gamma = \varphi_*(\Lambda) d(\varphi_*H) .$$
We conclude that any non--canonical transformation which is a symmetry for $\Gamma$, produces alternative Hamiltonian descriptions.  In infinitesimal form: if $X$ is an infinitesimal symmetry for the dynamics $\Gamma$,  $[\Gamma, X] = 0$, we get:
$$ \mathcal{L}_X(i_\Gamma \omega) = d \mathcal{L}_XH ,$$
then
$$i_\Gamma (\mathcal{L}_X \omega ) = d(\mathcal{L}_XH) ,$$
and the previous equation provides  an alternative Hamiltonian description for $\Gamma$ if $\mathcal{L}_X\omega$ is nondegenerate.  In 
relation with alternative descriptions, we notice that  there are additional ways to 
generate alternative Hamiltonian descriptions.  Let us consider for instance a $(1,1)$--tensor $T$ such that:
\begin{equation}\label{invariantT} 
\mathcal{L}_\Gamma T = 0 .
\end{equation}
Let us define as before eq. (\ref{twisted_diff}), the twisted differential:
$$ d_T f (X) = df(TX) .$$
Then from any constant of the motion $F$ we will have the closed 2--form:
$$ \omega_{T,F} = d d_T F ,$$
now if $\omega_{T,F}$ is nondegenerate, it provides an alternative Hamiltonian description for $\Gamma$.  In fact notice that:
$$i_\Gamma \omega_{T,F} = \mathcal{L}_{\Gamma} d_TF - d(i_\Gamma d_TF) ,$$
but $\mathcal{L}_\Gamma d_T F = 0$ because of the invariance condition (\ref{invariantT}) and the fact that $F$ is a constant of the motion.  Hence $-dF(T\Gamma)$ is a Hamiltonian function for $\Gamma$ with respect to the symplectic structure $\omega_{T,F}$.

\subsection{Integrability and normal forms}
Because of the previous discussion we should expect, therefore, that for vector fields with a large group of symmetries we will find always alternative Hamiltonian descriptions.   However we should stress that there are alternative Hamiltonian descriptions  which are not generated by diffeomorphisms.   

In conclusion, we should accept any diffeomorphism to reduce a vector field $\Gamma$ to its normal form.
Thus, if the orbit through $\Gamma$ contains a completely integrable system we can concentrate our attention on
the standard normal form we are familiar with when we construct action--angle variables, i.e., we could consider the form:
$$ \Gamma = \nu_j(I) \frac{\partial}{\partial \varphi_j} .$$
We should therefore study a normal form for integrable systems as emerging from the following conditions:

\begin{definition}\label{normal}  Given a vector field $\Gamma$ we will say that $\Gamma = \nu^j(f_1, \ldots, f_n) X_j$ is a normal form for $\Gamma$ (and that $\Gamma$ is integrable) if

\begin{itemize}
\item[i.] There exist $n$ functionally independent first integrals $f_1, \ldots, f_n$, such that $df_1 \wedge \cdots df_n  \neq 0$.

\item[ii.]  There exist $n$ complete vector fields $X_1, \ldots, X_n$ pairwise commuting $[X_j,X_l] = 0$ and independent $X_1 \wedge \cdots X_n \neq 0$, and

\item[iii.]  $\mathcal{L}_{X_j} f_l = 0$ for all $j,l = 1, \ldots, n$.

\end{itemize}
\end{definition}
We should notice that we have dropped the requirement for $\Gamma$ to be Hamiltonian and consequently that the ``frequencies'' $\nu^j$ are derivatives of the Hamiltonian function.

The usual Liouville--Arnold's theorem becomes now a particular way to find functions (coordinates) which reduce $\Gamma$ to normal form.

\medskip

A few remarks are in order here.

\begin{enumerate}
\item All integrable systems have the same normal form, then what distinguishes one system from another if any such distinction exists?

\item Which aspects of the normal form above for a given integrable system are able to discriminate integrable from superintegrable systems?
\end{enumerate}

In connection with the first query, we immediately notice that many interesting aspects on the qualitative structure of the orbits of the system are to be extracted from the normal form because we know that usually specific  integrable systems need not  be diffeomorphic among them.   We may rephrase our questions by investigating how many different orbits exist in $\mathfrak{X}(M)$ under the diffeomorphism group when each orbit is required to contain at least one element which is completely integrable.

To have an idea of the variety of situations we might be facing we shall investigate some variations on the theme of harmonic oscillators.

\subsection{Hamiltonian linear systems}  

We may consider as a particular instance of the analysis performed in the previous sections the isotropic harmonic oscillator.  
Thus, let us consider $M = \mathbb{R}^{2n} = \mathbb{C}^n$ and the linear system $\Gamma$ defined in cartesian coordinates $(x_j,y_j)$ by:
$$ \frac{d}{dt} x_j = \omega y_j, \quad \frac{d}{dt} y_j = -\omega x_j ,$$
then,
$$ \Gamma = \sum_j \omega \left( y_j \frac{\partial}{\partial x_j}  - x_j \frac{\partial}{\partial y_j}\right)  .$$
Introducing complex coordinates $z_j = y_j + ix_j$, we obtain:
$$ \frac{d}{dt} z_j = i\omega z_j, \quad \frac{d}{dt} \bar{z}_j = -i\omega \bar{z}_j ,$$
The algebra of first integrals is generated by the quadratic forms $z_l\bar{z}_m$ and because $\Gamma$ is proportional to the linear vector field
defined by the complex structure on $\mathbb{C}^n$, we conclude that its group of linear symmetries is $GL(n,\mathbb{C})$.    For a
given factorization $\Gamma = \Lambda (dH)$, the homogeneous space $GL(n,\mathbb{C})/ GL(n,\mathbb{C})\cap \mathrm{Sp}(2n, \mathbb{R})$ parametrizes alternative Hamiltonian descriptions, however not all of them.   

What will happen then for a generic linear sytems?  Given a generic linear system, represented by the matrix $A$, it has a decomposition:
\begin{equation}\label{factorization}
A =\Lambda \cdot H ,
\end{equation}
with $\Lambda$ a nondegenerate skew--symmetric matrix and $H$ a symmetric matrix if and only if $\mathrm{Tr} A^{2k+1} = 0$, $k = 0,1,2,\ldots $.   The proof is easily given in one direction.  Indeed, $A^T = -H\Lambda$, and therefore all odd powers must have the same property, implying the assertion.

For nongeneric matrices conditions guaranteeing the existence of the factorization (\ref{factorization})  are more cumbersome and we refer to the paper \cite{Gi98} for a full discussion.

If $T$ is a linear transformation such that $TAT^{-1} = A$, we have:
$$ A = (T\Lambda T^{t})((T^{-1})^tHT^{-1}) ,$$
providing an alternative decomposition of $A$ if $T$ is not canonical.

For instance, when $A$ is generic all symmetries are generated by powers of $A$.
The non--canonical ones are given by even powers, therefore:
$$  T = e^{\lambda A^{2k}}, \quad k = 1,2, \ldots ,$$
will be a non--canonical symmetry for any value of $\lambda$.

For instance, two alternative descriptions for the isotropic harmonic oscillator in $\mathbb{R}^4$ are given by:
\begin{eqnarray*} \Lambda_1 &=& \frac{\partial}{\partial p_1} \wedge \frac{\partial}{\partial q_1} + \frac{\partial}{\partial p_2} \wedge \frac{\partial}{\partial q_2}, \quad  H_1 = \frac{1}{2} \omega (p_1^2+ p_2^2 + q_1^2 + q_2^2) , \\
\Lambda_2 &=& \frac{\partial}{\partial p_1} \wedge \frac{\partial}{\partial q_2} + \frac{\partial}{\partial p_2} \wedge \frac{\partial}{\partial q_1}, \quad  H_2 = \omega (p_1 p_2 + q_1 q_2) .
\end{eqnarray*}

A particular  invariant $(1,1)$--tensor field is defined by:
$$T = dq_1 \otimes \frac{\partial}{\partial q_2} + dq_2 \otimes \frac{\partial}{\partial q_1} + dp_1 \otimes \frac{\partial}{\partial p_2} + dp_2 \otimes \frac{\partial}{\partial p_1} .$$
We may then consider the 2--form $dd_TF$ with $F = \frac{1}{4}(p_1^2+ p_2^2+ q_1^2 + q_2^2)^2$ and get:
\begin{eqnarray*} dd_TF &=& d(p_1^2+ p_2^2+ q_1^2 + q_2^2) \wedge d(p_1 p_2 + q_1 q_2) + \\
&& + 2(p_1^2+ p_2^2+ q_1^2 + q_2^2) (dq_2\wedge dq_1 + dp_1 \wedge dp_2) .
\end{eqnarray*}

We finally remark that the selection of a specific decomposition with the Hamiltonian being positive definite gives a group of canonical symmetries which is the unitary group:
$$ GL(n, \mathbb{C}) \cap O(2n, \mathbb{R}) = U(n) ,$$
therefore the system may be thought of as a ``quantum--like'' system  \cite{Er10}.
  
\section{The group of diffeomorphisms of an integrable system}

If $\Gamma$ is an integrable system possessing a normal form like in Def. \ref{normal}, then because $X_1, \ldots , X_n$ are complete and pairwise commuting we can define an action of the Abelian group $\mathbb{R}^n$ onto $M$.    Moreover, because $\mathcal{L}_{X_j} \nu^k = 0$, we could redefine the vector field $X_j$ to $Y_j = \nu^j X_j$ (no summation on $j$) and we would still have pairwise commuting vector fields $[Y_j, Y_k ] = 0$ for all $j,k = 1, \ldots, n$.   Notice that the completeness condition will not be spoiled by redefining the vector fields in this form, i.e., the vector fields $Y_k$ will be complete (but with a different parametrization) and we would define an alternative action of $\mathbb{R}^n$ on $M$.

Thus what seems to matter is not the particular action of $\mathbb{R}^n$ but rather the integral leaves of the involutive distribution generated by $X_1, \ldots, X_n$.    In those cases where the leaves are compact, say tori, we could require the choice of an action of $\mathbb{R}^n$ that factors to an action of $\mathbb{T}^n = \mathbb{R}^n / \mathbb{Z}^n$.  Moreover we could select a particular basis of vector fields such that $X_1, \ldots, X_n$ each generate the action of a closed subgroup.   With these particular prescriptions we decompose our vector field $\Gamma$.  Let us denote the particularly chosen generators of closed subgroups by $Z_1, \ldots, Z_n$, then:
$$ \Gamma = \omega^j Z_j .$$

Thus when the closure of generic orbits of $\Gamma$ are $n$--dimensional, the system does not have additional first integrals.  If for some leaves the closure of the orbits does not coincide with the full torus, there are additional invariant relations.  When for any initial conditon, the closure of the orbits is some $k < n$ dimensional torus then the system has additional first integrals and it is superintegrable.   

When the closure is one dimensional for all initial conditions, the system is maximally superintegrable.   

It is now clear that the geometry of integrable and superintegrable systems is associated with a toroidal generalized bundle, i.e., projections on the manifold $M$ which  have fibers diffeomorphic to tori of dimension going from one to $n$.  This is exactly the situation that happens if $M$ is a compact $2n$--dimensional symplectic manifold and the dynamics is invariant under the action of a torus group $\mathbb{T}^n$.  Then because of Atiyah's convexity theorem, the momentum map $J \colon M \to \mathfrak{t}^*$ is a surjective map onto a convex polytope $P\subset \mathfrak{t}^*$.   The fibers of $J$ are invariant tori that on points on the interior of $P$ are $n$-dimensional.  The fibers corresponding to the boundary of the polytope are lower dimensional tori \cite{At82}.  

Then the associated subgroup of diffeomorphisms of $M$ determined by $\Gamma$ is the subgroup of bundle automorphisms of such generalized toroidal bundle.  
As we will see later on, in connection with specific integrable or superintegrable systems, the most important obstruction to their being diffeomorphic is the energy--period theorem, that actually puts a restriction on the nature of the toroidal bundle of the system and in consequence on its group of diffeomorphisms.

\section{Oscillators and nonlinear oscillators}

To illustrate the previous situation we give now a few examples.

We may consider the isotropic Harmonic oscillator with two degrees of freedom.   Say $M = \mathbb{R}^4$, $\omega = \sum_a dp_a \wedge dq_a $, and  $H_0 = \frac{1}{2} \sum_a (p_a^2 + q_a^2)$.    In this case on $\mathbb{R}_0^4 = \mathbb{R}^4 - \{ 0 \}$, an open dense submanifold, the dynamics generates orbits with coincide with their closure, they are one--dimensional.  
The toroidal bundle is provided by:

$$ S^1 \to \mathbb{R}_0^4 \to S^2 \times \mathbb{R}^+ .$$

We have $\mathbb{R}_0^4 \cong S^3 \times \mathbb{R}^+$ and the dynamics induces the Hopf fibration $S^1 \to S^3 \to S^2$.  The subgroup of diffeomorphisms of  $\mathbb{R}_0^4$ which is selected by the fibration is the group of projectable diffeomorphisms.   Clearly this large group of symmetries, when applied to a chosen Hamiltonian description will generate many more alternative descriptions. 

It should be remarked that the alternative Hamiltonian description provided by $dp_1 \wedge dq_1 - dp_2 \wedge dq_2 $, and
$$ H = \frac{1}{2} (p_1^2 + q_1^2) - \frac{1}{2} (p_2^2 + q_2^2) $$
cannot be derived from the standard one with the positive definite Hamiltonian $H_0$ because the diffeomorphism would preserve the signature of $H$  (because it cannot map compact energy levels, the ones of the Hamiltonian $H_0$, into non--compact ones, the ones of $H$).  

The system is actually superintegrable, indeed the quotient manifold under the action of the dynamics is three dimensional instead of two dimensional.

This example generalizes to any finite dimension and we have again for the $n$--dimensional isotropic harmonic oscillator the fibration:
$$ S^1 \to \mathbb{R}_0^{2n} \to \mathbb{C}P^{n-1} \times \mathbb{R}^+ .$$

Again the symmetry group for the dynamics is the group of diffeomorphisms projectable under the previous projection, hence it is diffeomorphic to the central extension of Diff$(\mathbb{C}P^{n-1}\times \mathbb{R}^+)$ by $U(1)$.

The one--form $\frac{1}{n} \sum_k \left( \frac{p_k dq_k - q_k dp_k}{\omega (p_k^2  + q_k ^2)} \right) = d\tau$ has the property that $i_\Gamma d\tau = 1$.  Any closed two--form on the quotient manifold and any function on the quotient manifold without critical points on the invariant open dense submanifold specified by $p_k^2 + q_k ^2 \neq 0$, $k = 1, 2, \ldots, n$, give rise to an alternative Hamiltonian desription.  

This example is a normal form for maximally superintegrable systems with one--dimensional  closed orbits and constant period. 

In higher dimensions we can consider the Hamiltonian $H = \sum_a \omega_a H_a$, with $H_a = \frac{1}{2}(p_a^2 + q_a ^2)$, $\omega_a\in \mathbb{R}$.   The subgroup associated with the Hamiltonian vector fields $\Gamma_a$, $i_{\Gamma_a} \omega = dH_a$, are closed subgroups.  When all frequencies are pairwise irrational, i.e., $\omega_a/\omega_b$ is irrational, the closure of the generic orbit of $\Gamma$ is the full torus, in this case there cannot be additional constants of the motion.    When some of the frequencies are pairwise rational, the closure of a generic orbit is a torus of lower dimensions.  A particular example where the closure of the orbits goes from a one--dimensional torus to an $n$--dimensional one, depending on the initial conditioins, is provided by:

$$ H = \sum_a \pm (H_a)^2 .$$

In this case we may also find invariant relations for particular values of the initial conditions. 

This example gives rise to the so called nonlinear oscillators and has been considered in quantum mechanics to give interesting consequences at the level of Planck's distribution law and for alternative commutation relations \cite{Ma97}, \cite{Lo97}.

\section{Obstructions to the equivalence of integrable systems}\label{energy_period}

If we inquire about the obstructions to the existence of diffeomorphisms conjugating two integrable systems, the energy--period theorem provides the first and most important one.  We refer to the literature  \cite{Ne02}, \cite{Go69} for various versions of this theorem, however a ``cheap'' proof may be given as follows.    On the carrier space of our dynamical system let us consider the two--form $dp_k\wedge dq^k - dH\wedge dt$.  Consider now the evolution $t \mapsto t + \tau$, $p_k (t) = p_k (t + \tau)$, $q^k(t) = q^k(t+\tau ) $ and $H(t) = H(t')$ for all $t' \in \mathbb{R}$.   Because the evolution is canonical, we find  $dH \wedge d\tau = 0$ on the submanifold on which the period is a differentiable function on $(q^k, p_k)$.  It follows that the period $\tau$ and the Hamiltonian $H$ are functionally dependent.

It is well--known that the period of a dynamical system is an attribute of the vector field which is invariant under diffeomorphisms.  It follows that if two Hamiltonian systems have different periods, they cannot be connected via diffeomorphisms.   For instance the isotropic Harmonic oscillator and the Kepler problem cannot be connected by a diffeomorphism.   Indeed the map connecting solutions of the Harmonic oscillator and those of the Kepler problem, the Kustaanhneimo--Stiefel map, is a map defined on each energy level, i.e., for those orbits that have all the same period.  The map changes from one energy--level to another (see for instance \cite{Av05}, \cite{Av05b}).

A more simple example is provided by $H = \frac{1}{2} (p^2 + q^2)$ and $H' = (p^2 + q^2)^2$.  For the first one, the frequency is independent from the energy, while for the second one it depends on the initial conditions.    The two systems cannot be diffeomorphic.    This circumstance was the main motivation to introduce the classification of dynamical systems up to parametrization, i.e., up to conformal factors \cite{Ib94}.   

We hope we have made clear that the ``geometrical picture'' we have derived is the best one can do because each individual integrable system will give rise to infinitely many different situations which cannot be classified in a meaningful way otherwise (i.e., identifying a finite or a countable family of equivalence classes).

\bigskip

\paragraph{Acknowledgments}
This work was partially supported by MEC grant
MTM2010-21186-C02-02. and QUITEMAD programme.  G.M. would like to acknowledge the support provided by the Santander/UCIIIM Chair of Excellence programme 2011-2012.

\end{document}